# Similarity based Dynamic Web Data Extraction and Integration System from Search Engine Result Pages for Web Content Mining

Srikantaiah K C[1], Suraj M[1], Venugopal K R[1], and L M Patnaik[2]
[1] Department of Computer Science and Engineering,
University Visvesvaraya College of Engineering, Bangalore University, Bangalore , India
srikantaiahkc@gmail.com
[2] Honorary Professor, Indian Institute of Science, Bangalore, India

*Abstract*— **There is an explosive growth of information in the World Wide Web thus posing a challenge to Web users to extract essential knowledge from the Web. Search engines help us to narrow down the search in the form of Search Engine Result Pages (SERP). Web Content Mining is one of the techniques that help users to extract useful information from these SERPs. In this paper, we propose two similarity based mechanisms; WDES, to extract desired SERPs and store them in the local depository for offline browsing and WDICS, to integrate the requested contents and enable the user to perform the intended analysis and extract the desired information. Our experimental results show that WDES and WDICS outperform DEPTA [1] in terms of Precision and Recall.**

*Index Terms*— **Offline Browsing, Web Data Extraction, Web Data Integration, World Wide Web, Web Wrapper**

## I. Introduction

The World Wide Web (WWW) has now become the largest knowledge base in the human history. The Web encourages decentralized authorizing in which users can create or modify documents locally, which makes information publishing more convenient and faster than ever. Because of these characteristics, the Internet has grown rapidly, which creates a new and huge media for information sharing and exchange. Most of the information on the Internet cannot be directly accessed *via* the static link, must use Keywords and Search Engine. Web search engines are programs used to search information on the WWW and FTP servers and to check the accuracy of the data automatically. When searching for a topic in the WWW, it returns many links or web sites related *i.e.,* Search Engine Result Pages (SERP) on the browser to a given topic. Some data in the internet is visible to search engine is called surface web, where as some data such as dynamic data in dynamic database is invisible to search engine is called deep web.

There are situations in which the user needs those web pages on the Internet to be available offline for convenience. The reason being offline availability of data, limited download slots, storing data for future use, *etc.*. This essentially leads to downloading raw data from the web pages on the Internet which is a major set of the inputs to a variety of software that are available today for the purpose of data mining. Web data Extraction is the process of extracting the information that users are interested in, from Semi-structured or unstructured web pages and saving the information as the XML document or relationship model. During Web data extraction phase Search Engine Result Pages are crawled and stored in the local repository. Web database Integration is a process of extracting the required data from the web pages stored in the Local repository and Integrates the extracted data and stored in the database.

### A. Motivation

In the recent years there has been lot of improvements on technology with products differing in the slightest of terms. Every product needs to be tested thoroughly and internet plays a vital role in gathering information for the effective analysis of the products. In our approach, we replicate search engine result pages locally based on comparing page URLs with a predefined threshold. The replication is such that the pages are accessible locally in the same manner as on the web. In order to make the data available locally to the user for analysis we extract and integrate the data based on the prerequisites which are defined in the configuration file.

### B. Contribution

In a given set of web pages, it is difficult to extract matching data. So, we have to develop a tool that is capable of extracting the exact data from the web pages. In this paper, we have developed WDES algorithm, which provides offline browsing of the pages. Here, we integrate the downloaded content onto a defined database and provide a platform for efficient mining of the data required.

### C. Organization

The rest of this paper is organized as follows: Section II describes algorithms related to web data extraction, integration and crawling, Section III defines the problem, describes mathematical model and algorithm, Section IV describes the system architecture, Section V comprises of experimental results and analysis. The concluding remarks are summarized in section VI.

## II. RELATED WORK

Zhai et al., [1] propose an algorithm DEPTA for the structured data extraction from the web based on partial tree alignment, studying the problem structured data extraction from arbitrary web pages. The main objective is to automatically segment data records in a page, extract data items/fields from these records and store the extracted data in a database. It consists of two steps, *i.e* identifying individual records in a page and aligning and extracting data items from the identified records, using visual information and tree matching and a novel partial alignment technique respectively.

Ananthanarayanan et al., [2] propose a method for offline web browsing that is minimally dependent on the real time network availability. The approach defined is to make use of the Really Simple Syndication (RSS) feeds from web servers and pre-fetch all new content specified, defining the content section of the home page. It features intelligent pre-fetching, robust and resilient measures for intermittent network handling, template identifier and local stitching of the dynamic content into the template. It does not provide all the information in the page. Also, the content defined in the RSS feeds may not be updated nor they do provide for dynamic changes in the page.

Myllymaki et al., [3] describe ANDES, a software framework that provides a platform for building a production-quality Web data extraction process. Key aspects are that it uses XML technology for data extraction, including XHTML and XSLT and provides access to the deep Web. It addresses the issues of website navigation, data extraction, structure synthesis, data mapping and data integration. The framework shows that production-quality web data extraction is quite feasible and that incorporating domain knowledge into the data extraction process can be effective in ensuring the high quality of extracted data. Data validation technique, a cross system support, and a well established framework that could easily be made use of by any application are not discussed.

Yang et al., [4] propose a novel model to extract data from deep web pages. It four layers, among which the access schedule, extraction layer and data cleaner are based on the rules of structure, logic and application. The model first uses the two regularities of the domain knowledge and interface similarity to assign the tasks that are proposed from users and chooses the most effective set of sites to visit. Then, the model extracts and corrects the data based on logical rules and structural characteristics of acquired pages. Finally, it cleans and orders the raw data set to adapt the customs of application layer for later use by its interface.

Yin et al., [5] propose web page templates and DOM technology to effectively extract simple structured information from the web. The main contents include the methods based on edit distance, DOM document similarity judgement, clustering methods of web page templates and programming an information extraction engine. The method provides on the steps for information extraction DOM tree parsing and on how a page similarity judgement is to be made. The template extraction and reconstruction is depicted in an order of how the data is been parsed on the web page and in reconstructing the page to overcome the noise in the page. It does not parse through the dynamic content of scripts on the page.

Liu et al., [6] propose a method to extract main text from the body of a page based on the position of DIV. It reconstructs and analyzes DIV in a web page by completely simulating its display in browser, without additional information such as web page template and its implementation complexity is quite low. The core idea includes the concept of atomic DIV, *i.e.,* a DIV block that does not include other DIVs. Then filter out on redundant, reconstruct, filter invalids, clustering, reposition analysis and finally storing the elements of data in an array. The selected DIVs in selected array contain the main text of the page. We can get the main text by combining these DIVs. The method has a high versatility and accuracy. The majority of the web page content is been made up of tables and this approach does not address the table layout data. This drastically reduces the accuracy of the entire system.

Dalvi et al., [7] explore a novel approach based on temporal snapshots of web pages to develop a tree-edit model of HTML and use this model to improve wrapper construction. The model is attractive in that the probability that a source tree has evolved into a target tree can be estimated efficiently, in quadratic time in the size of the trees, making it a potentially useful tool for a variety of tree-evolution problems. An improvement on the robustness and performance and ways to prune the trees without hampering model quality is to be dealt with.

Novotny et al., [8] represent a chain of techniques for extraction of object attribute data from web pages. They discover data regions containing multiple data records and also provide for a detailed algorithm for detail page extraction based on the comparison of two html sub trees. They describe the extraction from two kinds of web pages: master pages containing structured data about multiple objects and detail pages containing data about single product respectively. They combine the techniques of the master page extraction algorithm, detail page extraction algorithm and comparison of sources of two web pages. The approach makes use of the selection of attribute values based on the Document Object Model (DOM) structure described in the web pages. It has better precision of extraction of values from pages defined in the master and detailed format and also having to minimize the user effort to the core. Enhances to the approach may include the page level complexity of having multiple interleaved detail pages to be traversed, coagulation of different segments in the master page and series implementation of pages.

Nie et al., [9] provide an approach of obtaining data from the result set pages by crawling through the pages for data extraction based on the classification of the URL (Unified Resource Locator). It extracts data from the pages by computing the similarity between the URL's of hyperlinks and classifying them into four categories, where each category maps to a set of similar web pages, which separate result pages from others. Then makes use of the page probing method to verify the correctness of classification and improve the accuracy of crawled pages. The approach makes use of the minimum edit distance algorithm and URL-field algorithm to calculate the similarity between URLs of hyperlinks respectively. However there are a few constraints to this approach. It is not able to resolve issues of pages related to the partial page refreshments by the use of javascript engines.

Papadakis et al., [10] describe STAVIES, a novel system for information extraction from web data source through automatic wrapper generation using clustering technique. The approach is to exploit the format of the Web pages to discover the underlying structure in order to finally infer and extract pieces of information from the web page. The system can operate without human intervention and does not require any training.

Chang et al., [11] survey the major web data extraction systems and compare them in three dimensions: the task domain, the automation degree and the techniques used. These approaches emphasize on availability of robust, flexible Information Extraction (IE) systems that transform the web pages into program-friendly structures such as a relational database becomes a great necessity. The paper mainly focuses on the IE from semi structured documents and discusses only those that have been used for web data. Based on the survey it makes many points such as the trend of developing highly automatic IE systems, which saves not only the effort for programming, but also the effort for labeling, enhancements for applying the techniques to non-html documents such as medical records and curriculum vitae to facilitate the maintenance of larger semi structured documents.

Angkawattanawit et al., [12] propose an algorithm to improve harvest rate by utilizing several databases like seed URLs, topic keywords and URL relevance predictors that are built from previous crawl logs. Seed URLs are computed using BHITS [13] algorithm on previously found pages by selecting pages with high hub and authority scores that will be used for future recrawls. The interested Keywords for the target topic are extracted from anchor tags and title of previously found relevant pages. Link crawl priority is computed as a weighted combination of popularity of the source page, similarity of link anchor text to topic keywords and predicted link score which is based on previously seen relevance for that specific URL.

Aggarwal et al., [14] propose an approach to crawl the interested web pages using the concept of "intelligent crawling". In this concept the user can specify an arbitrary predicate such as keywords, document similarity, *etc.*, which are used to determine documents relevance to the crawl and the system adapts itself in order to maximize the harvest rate. A probabilistic model for URL priority prediction is trained using URL tokens, information about content of in-linking pages, number of sibling pages matching the predicate so far and short-range locality information

Chakrabarti et al., [15] propose models for finding URL visit priorities and page relevance. The model for URL ranking called "apprentice" is on-line trained by samples consisting of source page features and the relevance of the target page but the model for evaluating page relevance can be anything that outputs a binary classification. For each retrieved page, the apprentice is trained on information from baseline classifier and features around the link extracted from the parent page to predict the relevance of the page pointed to by the link. Those predictions are then used to order URLs in the crawl priority queue. Number of false positives has decreased significantly.

Ehrig et al., [16] propose an ontology-based algorithm for page relevance computation which is used for web data extraction. After preprocessing, words occurring in the ontology are extracted from the page and counted. Relevance of the page with regard to user selected entities of interest is then computed by using several measures such as direct match, taxonomic and more complex relationships on ontology graph. The harvest rate of this approach is better than baseline focused crawler.

Srikantaiah et al., [17] propose an algorithm for web data extraction and integration based on URL similarity and Cosine Similarity. Extraction algorithm is used to crawl the relevant pages and stores in local repository. Integration algorithm is used to integrate the similar data in various records based on cosine similarity.

III. PROPOSED MODEL AND ALGORITHMS

*A. Problem Definition*

Given a start page URL and a configuration file, the main objective is to extract pages which are hyperlinked from the start page and integrate the required data for analysis using data mining techniques. The user has sufficient space on the machine to store the data that is downloaded. The basic notations used in the model are shown in Table 1.

TABLE 1
BASIC NOTATIONS

| | |
|---|---|
| $S$ | : Start Page URL |
| $C, C_i$ | : Configuration File |
| $l$ | : Depth of Recursion |
| $W$ | : Set of Search Engine Result Pages |
| $H(W)$ | : Hyperlinks Set of W |
| $Cl$ | : Current Level |
| $Lp$ | : Local Path to hyperlinks |
| $T_o$ | : Threshold for Similarity |

*B. Mathematical Model*

*Web Data Extraction using Similarity Function (WDES)*: A connection is been established to the given URL *S* and the page is processed with the parameters obtained from the configuration file *C*. On completion of this, we obtain the web document that contains the links to all the desired contents that are obtained out of the search performed. The web document contains individual sets of links that are displayed on each of the search results pages that are obtained. For example, if a search result obtained contains 150 records displayed as 10 records per page (in total 15 pages of information), we would have 15 sets of web documents each containing 10 hyperlinks pointing to the required data. This forms the set of web documents, *W*. i.e.,

$$W = \{w_i : 1 \leq i \leq n\}. \quad (1)$$

Each web document $w_i \in W$ is read through to collect the hyperlinks that are contained in it, that are to be fetched to obtain the data values. We, represent this hyperlink set as *H(W)*. Thus, we consider *H(W)* as a whole set containing all the sets of hyperlinks on each page $w_i \in W$. i.e.,

$$H(W) = \{H(w_i) : 1 \leq i \leq n\}. \quad (2)$$

Then, considering each hyperlink $h_j \in H(w_i)$, we find the similarity between $h_j$ and *S*, using (3)

$$SIM(h_j, S) = \frac{\sum_{i=1}^{\min(nf(h_j), nf(S))} fsim(f_i h_j, f_i S)}{(nf(h_j) + nf(S))/2} \quad (3)$$

where *nf(X)* is the number of fields in *X* and *fsim(f_i h_j, f_i S)* is defined as

$$fsim(f_i h_j, f_i S) = \begin{cases} 1 & if\ f_i h_j = f_i S \\ 0 & if\ f_i h_j \neq f_i S \end{cases}. \quad (4)$$

The similarity *SIM(h_j, S)* is the value that lies between 0 and 1, this value is used to compare with the defined threshold $T_o$ (0.25), we download the page corresponding to $h_j$ to local repository if $SIM(h_j, S) \geq T_o$. The detailed algorithm of WDES is given in Table 2.

The algorithm WDES navigates the search result page from the given URL *S* and configuration file *C* and generates a set of web documents *W*. Next, call the function *Hypcollection* to collect hyperlinks of all pages in $w_i$, indexed by $H(w_i)$, page corresponding to $H(w_i)$ is stored in the local repository. The function *webextract* is recursively called for each $H(w_i)$. Then, for each $h_i \in H(w_i)$, similarity between $h_i$ and *S* is calculated using (3), if $SIM(h_i, S)$ is greater than the threshold $T_o$, then page corresponding to $h_i$ is stored and collect all the hyperlinks in $h_i$ to *X*. Continue this process for *X*, until it reaches maximum depth *l*.

*Web Data Integration using Cosine Similarity(WDICS)*: The aim of this algorithm is to extract data from the downloaded web pages (those web pages that are available in the local repository *i.e.,* output of WDES algorithm) into the database based on attributes and keywords from the configuration file $C_i$. We collect all result pages *W* from local repository indexed by *S*, then *H(W)* is obtained by collecting all hyperlinks from *W*, considering each hyperlink $h_j \in H(w_i)$ such that $k \in$ keywords in $C_i$. On existence of *k* in $h_j$, we populate the new record set *N[m, n]* by passing page $h_j$ and obtaining values defined with respect to the *attributes[n]* in $C_i$. We then populate the old record set *O[m, n]* by obtaining all values with respect to *attributes[n]* in database. For each record *i*, $1 \leq i \leq m$ we find the similarity between *N[i]* and *O[i]* using cosine similarity,

$$Sim\,Record(N_i, O_i) = \frac{\sum_{j=1}^{n} N_{ij}O_{ij}}{\sqrt{\sum_{j=1}^{n} N^2_{ij} \sum_{j=1}^{n} O^2_{ij}}} \quad (5)$$

If similarity between records is equal to zero, then we compare each *attribute[j]* $1 \leq j \leq n$ in the records and form *IntegratedData* with use of Union operation and store in the database.

$$IntegratedData = Union(N_{ij}, O_{ij}). \quad (6)$$

The detailed algorithm of WDICS is shown in Table 3. The algorithms *WDES* and *WDICS* respectively extract and integrate data in Depth First Search (DFS) manner. Hence their complexity is $O(n^2)$, where n is the number of hyperlinks in H(W).

TABLE 2
ALGORITHM: WEB DATA EXTRACTION USING SIMILARITY

```
Input
    S : Starting Page URL.
    C : Parameter Configuration File.
    l : Level of Data Extraction.
    T_o : Threshold.
Output: Set of Webpages in Local
    Repository.
begin
    W=Navigate to Web document on Given
        S and automate page with C
    H(W)=Call: Hypcollection(W)
    for each H(w_i) ∈ H(w)
        Save page H(w_i) on local Machine
            page P
        Call: Webextract(H(w_i),0,pageppath)
    end for
end
Function Hypcollection(W)
begin
    for each w_i ∈ W do
        H(w_i)=Collect all hyperlinks in w_i
    end for
    return H(W)
end
Function Webextract(Z, cl, lp)
Input
    Z : set of URLs.
    cl : Current level.
    lp : local path to Z.
Output: Set of Webpages in Local
    Repository.
begin
    for each h_i ∈ Z do
        if SIM(h_i, S) ≥ T_o then
            Save h_i to Fh_i
            X=collect URLs from h_i and change
                its path in lp
            if( cl < l)
                Call: Webextract(X,cl + 1,
                    pageppath of X)
            end if
        end if
    end for
end
```
FUNCTION (WDES)

TABLE 3
ALGORITHM: WEB DATA INTEGRATION USING COSINE SIMILARITY
(WDICS)

```
Input
 S : Starting Page URL stored in local repository (output of
    WDES).
 C_i : Configuration File (Attributes and Keywords).
Output: Integrated Data in Local Repository.
Begin
  H(w)=Call: Hypcollection(S)
  for each H(w_i) ∈ H(w) do
    Call: Integrate(H(w_i))
  end for
end
Function Integrate(X)
Input: X : set of URLs.
Output: Integration of Values of Attributes Local Repository.
begin
  for each hi ∈ Z do
    if( h_i contain keyword) then
      new[m][n]=parse page to obtain values of defined
           attributes[n] in C_i
      old[m][n]=obtain all values of attributes[n] from
           repository
      for each record i do
        if(SimRecord(new[i], old[i])==1) Skip
        end if
        else
        for each attribute j do
          if ( new[i][j] Not Equal to old[i][j] )
            IntegratedData=union(new[i][j],old[i][j])
          end if
        end for
        store IntegartedData in local repository
      end for
      X=collect all links for h_i
      if (X not equal to NULL) Call: Integrate(X)
    end if
  end for
end
```

## IV. SYSTEM ARCHITECTURE

The entire system has been divided into three blocks that are able to accomplish the dedicated tasks, working independently from one another. The first block namely, the Web Data Extractor connects to the internet to extract the web pages described by the user and stores it onto a local repository. Also, it stores the data in a format that is likely to be an offline browsing system. Offline Browsing means that the user is able to browse through the pages that are been downloaded, by the use of a web browser without having to connect to the internet. Thus, it would be convenient for the user to go through the pages as and when he needs.

The second block namely, the Database Integrator extracts the vital piece of information that the user needs from the downloaded web pages and creates a database with the set of tables and respective attributes in the database. These tables are populated with the data extracted from the locally downloaded web pages. This makes it easy for the user to query out his needs from the database.

The overall view of the system is as shown in Fig. 1. The system initializes on a set of inputs for each block. These blocks are highlighted and framed according to the flow of data. The inputs that are needed for the start of the first block, *i.e.,* the Web Data Extractor are fed in through the initialize phase. On receiving the inputs the search engine performs its task of navigating to the page on the given URL and entering the criteria defined on the configuration file. This will populate a page from which the actual download can start.

The process of extraction is to achieve the task of downloading the contents from the web and to store them onto the local repository. Then onwards the process of extracting the pages iterates over, resulting in the outcome of offline web pages (termed extracted pages). The extracted pages are available offline on a pre-described local repository. The pages in the repository can be browsed through in the same manner as that available on the internet with the help of a web browser. The noticeable thing here is that the pages are available offline, thereby are much faster to be accessed.

The outcome of first block is to be chained with other attributes that are essential for the second block namely, the Database Integrator. The task of second block is to extract the vital information from the extracted pages, process it, and store it in accordance onto the database. The database is pre-defined in the set of attributes, together with the tables. This block needs the presence of the usage attributes that are extracted from the downloaded web content. The usage attributes are mainly defined in a file termed as configuration file that the user needs to prepare before the execution. The file also contains the table to which the data extracted are to be added together with the table attributes and mapping.

On successful entries of the input, the integration block is able to accomplish the task of extracting the content from the web pages. Since, the web pages do tend to remain in the same format as on the internet, it is easy to be able to navigate across these pages as the links refer to the locally extracted pages. The extracted content is then processed to meet the desired data attributes that are listed on the database and the values are dumped onto it. This essentially creates rows of extracted information in the database. The outcome of second block results in tuples in the database that are essentially extracted out from the extracted pages got from first block.

Now that the database is been formed with the vital information contained in it, it would well be the task of the Analyzer, *i.e.,* the third block referred as GUI to provide the user with the functionality on how to deal with the contained data. The GUI provides for the interfaces that the user is able to achieve so as to obtain the data contained in the database, as and how he needs it. It acts as a miner that provides the user with the information obtained from the result of queries that are defined. The GUI also has options on referring the other blocks as requested by the user. This indeed interfaces all the three blocks, thereby providing the user with a better understanding and handling feature.

It is quite essential to know a few things that relate to the working of the entire system. First of all, it is very much essential that the inputs, initializations and the pre-requisite data such as the usage attributes and/or configuration files that are been defined, do fall in a particular order and stick to the conventions defined on them. It is important to have the blocks to perform the tasks in order. The next thing is that, although the blocks do persist to function independently it is important that without the essential requirements, it is of no use to extract its functionality. Unless the pre-requisites of the blocks are not been met the functionality that they accomplish is of no use. Similarly, to query out the needs of the user through the help of the GUI, it is essential that the data is available in the database. It is also essential that the first block and second block be run often and in unison so as to have an update on the current and on-going dataset.

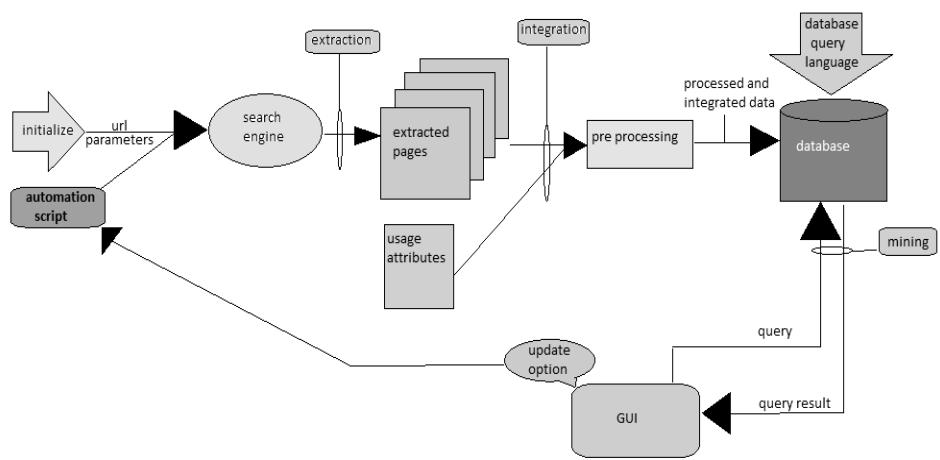

Figure 1. System Architecture

## V. Experimental Results

The algorithms are independent of usage and that they could be used on any given platform and dataset. The experiment was conducted on the Bluetooth SIG Website [18], The Bluetooth SIG (Special Interest Group) is a body that oversees the licensing and qualification of Bluetooth enabled product. It maintains the database of all the qualified products in a database which can be accessed through queries on its website and hence is a domain specific search engine. The qualification data contains a lot of parameters including company name, product name, any previous qualification used, *etc.*. The main tasks for the Bluetooth SIG are to publish Bluetooth specifications, administer the qualification program, protect the Bluetooth trademarks and evangelize Bluetooth wireless technology. The key idea behind the approach is to collect and automate the collection of competitive and market information from the Bluetooth SIG site.

The site contains data in the form of list that is displayed on the start-up page. The display is formed based on the three types of listings; PRD 1.0, PRD 2.0 and EPL. The PRD 1.0 and PRD 2.0 are the qualified products and design list and EPL are the end product list being displayed. Each of the PRD's listed here are products that contain the specifications involved in them. The EPL are those that are formed in unison of the PRD's. Each PRD is identified by a QDID (an id for uniqueness) and each EPL is identified by the Model. Each PRD may have many numbers of related EPL's and each EPL may have many numbers of related PRD's.

The listings as shown in Fig. 2, contain links which navigate to the detailed content of the products that are been displayed here. The navigations on the page reach to $N$ number of pages before the case of termination. Thereby we may want to parse across the links to reach all the places that the data we require is obtained. Here, although we have all the data being displayed on the web site, it is still not possible for us to prepare an analysis report based on the same. We want to play around with th data to get it to the form that we deserve it to be. Thereby, we incorporate the use of our algorithms to extract, integrate and mine the data from this site.

The experimental setup involves a Pentium Core 2 Duo machine with 2 GB RAM running windows. The algorithms have been implemented using a Java JDK and JRE 1.6, Java Parser with an access to a SQLite database and active broadband connection. Data has been collected from www.bluetooth.org, which lists the qualified products of Bluetooth devices. We have extracted the pages with dates ranging from October 2005 to June 2011, all of which make up 92 pages, with each page containing 200 records of information and data extraction was possible from each of these pages. Hence, we have a cumulative set of data for comparison based on the data extracted on the given attribute mentioned in the configuration file.

Precision is defined as the ratio of correct pages and extracted pages and recall is defined as the ratio of extracted pages and total number of pages. They are calculated based on the records extracted by our model, the records found by the search engine and the total available records in the Bluetooth website. For different attributes as shown in the Table 4, the Recall and Precision are calculated and their comparisons are as shown in Figures 3 and 4 respectively. It is observed that the Precision of WDICS increases by 4% and the Recall increases by 2% compared to DEPTA. Therefore, WDICS is more efficient than DEPTA because when an object is dissimilar to its neighbouring objects, DEPTA fails to identify all records correctly.

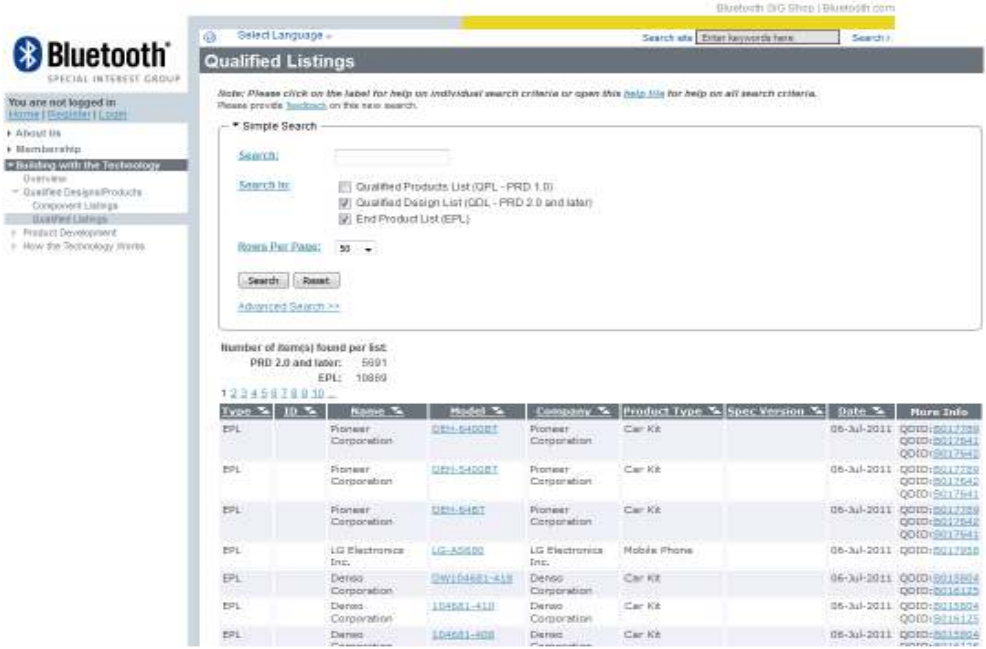

Figure 2. SERP in Bluetooth Sig Website

TABLE 4
PERFORMANCE EVALUATION BETWEEN WDICS AND DEPTA

| Attributes | Total Records(TR) | DEPTA | | WDICS | |
|---|---|---|---|---|---|
| | | Extracted Records(ER) | Correct Records(CR) | Extracted Records(ER) | Correct Records(CR) |
| Name | 18234 | 18204 | 17325 | 18234 | 18234 |
| Model | 18234 | 17860 | 17010 | 18060 | 18060 |
| Company | 18234 | 18095 | 17208 | 18234 | 18198 |
| Spec Version | 5508 | 5410 | 5016 | 5508 | 5426 |
| Product Type | 18234 | 17834 | 17015 | 18234 | 18045 |

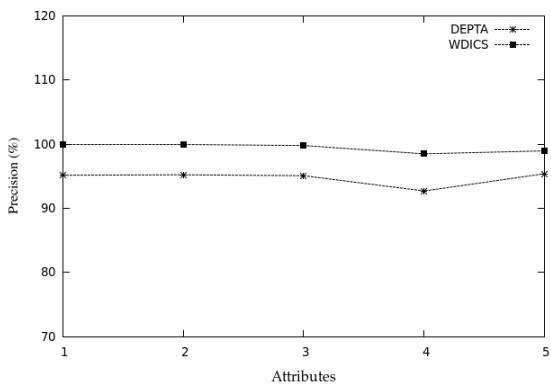

Figure 3. Precision Vs. Attrinbutes

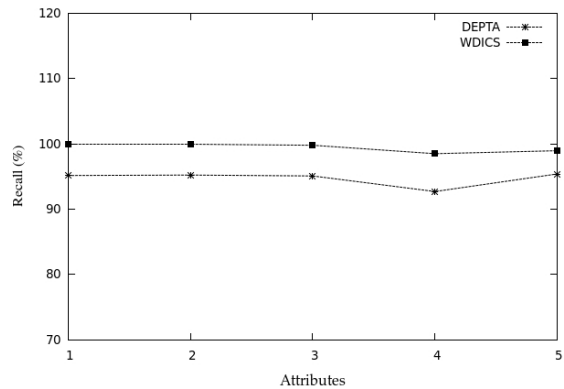

Figure 4. Recall Vs. Attributes

## VI. CONCLUSIONS

One of the major issues in Web Content Mining is the extraction of precise and meticulous information from the Web. In this paper, we propose two similarity based mechanisms; WDES, to extract desired SERPs and store them in the local depository for offline browsing and WDICS, to integrate the requested contents and enable the user to perform the intended analysis and extract the desired information. This results in faster data processing and effective offline browsing for saving time and resources. Our experimental results show that WDES and WDICS outperform DEPTA in terms of Precision and Recall. Further**,** different Web mining techniques such as classification and clustering can be associated with our approach to utilize the integrated results more efficiently.